\pdfoutput=1
\documentclass[12pt,a4paper]{article}

\usepackage{ifthen} 
\newboolean{pdflatex}
\setboolean{pdflatex}{true} 

\newboolean{articletitles}
\setboolean{articletitles}{true} 

\newboolean{uprightparticles}
\setboolean{uprightparticles}{false} 

\def\paperauthors{LHCb collaboration} 
\def\paperasciititle{Search for the rare decays W+- ->Ds gamma and Z->D0 gamma at LHCb} 
\def\papertitle{Search for the rare decays \WDspg and \ZDzDzbg at \lhcb} 

\def\WDsResults {{\ensuremath{6.5\times10^{-4}}}\xspace}
\def\WDsRelaResults {{\ensuremath{6.1\times10^{-3}}}\xspace}
\def\ZDzResults {{\ensuremath{2.1\times10^{-3}}}\xspace}
\def\ZDzRelaResults {{\ensuremath{6.4\times10^{-2}}}\xspace}
\def\lumi {{\ensuremath{2.0\invfb}}\xspace}

\def\paperkeywords{{High Energy Physics}, {LHCb}} 
\def\papercopyright{\the\year\ CERN for the benefit of the LHCb collaboration} 
\def\paperlicence{CC BY 4.0 licence}
\def\paperlicenceurl{https://creativecommons.org/licenses/by/4.0/}

\usepackage[top=1in, bottom=1.25in, left=1in, right=1in]{geometry}

\usepackage{microtype}
\usepackage{lineno}  
\usepackage{xspace} 
\usepackage{caption} 

\usepackage{graphicx}  
\usepackage{color}
\usepackage{colortbl}
\graphicspath{{./figs/}} 

\usepackage{amsmath} 
\usepackage{amssymb}
\usepackage{amsfonts}
\usepackage{upgreek} 

\newcommand*\patchAmsMathEnvironmentForLineno[1]{%
\expandafter\let\csname old#1\expandafter\endcsname\csname #1\endcsname
\expandafter\let\csname oldend#1\expandafter\endcsname\csname
end#1\endcsname
 \renewenvironment{#1}%
   {\linenomath\csname old#1\endcsname}%
   {\csname oldend#1\endcsname\endlinenomath}%
}
\newcommand*\patchBothAmsMathEnvironmentsForLineno[1]{%
  \patchAmsMathEnvironmentForLineno{#1}%
  \patchAmsMathEnvironmentForLineno{#1*}%
}
\AtBeginDocument{%
\patchBothAmsMathEnvironmentsForLineno{equation}%
\patchBothAmsMathEnvironmentsForLineno{align}%
\patchBothAmsMathEnvironmentsForLineno{flalign}%
\patchBothAmsMathEnvironmentsForLineno{alignat}%
\patchBothAmsMathEnvironmentsForLineno{gather}%
\patchBothAmsMathEnvironmentsForLineno{multline}%
\patchBothAmsMathEnvironmentsForLineno{eqnarray}%
}

\usepackage{hyperxmp}

\usepackage[pdftex,
            pdfauthor={\paperauthors},
            pdftitle={\paperasciititle},
            pdfkeywords={\paperkeywords},
            pdfcopyright={Copyright (C) \papercopyright},
            pdflicenseurl={\paperlicenceurl}]{hyperref}

\usepackage[colorinlistoftodos,textsize=scriptsize]{todonotes}

\usepackage[bottom,flushmargin,hang,multiple]{footmisc}

\usepackage[all]{hypcap} 

\usepackage{xspace} 
\usepackage{upgreek}

\def\lhcb   {\mbox{LHCb}\xspace}
\def\atlas  {\mbox{ATLAS}\xspace}
\def\cms    {\mbox{CMS}\xspace}

\def\cdf    {\mbox{CDF}\xspace}

\def\MagUp {\mbox{\em Mag\kern -0.05em Up}\xspace}

\ifthenelse{\boolean{uprightparticles}}%
{
 
 \def\Pgamma      {\ensuremath{\upgamma}\xspace}

 \def\Pmu         {\ensuremath{\upmu}\xspace}                 
 \def\Pnu         {\ensuremath{\upnu}\xspace}                 
                  
 \def\Ppi         {\ensuremath{\uppi}\xspace}

 \def\Ptau        {\ensuremath{\uptau}\xspace}

 \def\PDelta      {\ensuremath{\Delta}\xspace}                 
 \def\PXi         {\ensuremath{\Xi}\xspace}                 
 \def\PLambda     {\ensuremath{\Lambda}\xspace}                 
 \def\PSigma      {\ensuremath{\Sigma}\xspace}                 
 \def\POmega      {\ensuremath{\Omega}\xspace}                 
 \def\PUpsilon    {\ensuremath{\Upsilon}\xspace}
 \let\oldPi\Pi
 \def\PPi         {\ensuremath{\oldPi}\xspace}

 \def\PB      {\ensuremath{\mathrm{B}}\xspace}                 
                  
 \def\PD      {\ensuremath{\mathrm{D}}\xspace}

 \def\PK      {\ensuremath{\mathrm{K}}\xspace}

 \def\PW      {\ensuremath{\mathrm{W}}\xspace}

 \def\PZ      {\ensuremath{\mathrm{Z}}\xspace}                 
                  
 \def\Pb      {\ensuremath{\mathrm{b}}\xspace}                 
 \def\Pc      {\ensuremath{\mathrm{c}}\xspace}

 \def\Pi      {\ensuremath{\mathrm{i}}\xspace}

 \def\Ps      {\ensuremath{\mathrm{s}}\xspace}

 \def\thebaroffset{0.0em}
}
{
 
 \def\Pgamma      {\ensuremath{\gamma}\xspace}

 \def\Pmu         {\ensuremath{\mu}\xspace}                 
 \def\Pnu         {\ensuremath{\nu}\xspace}                 
                  
 \def\Ppi         {\ensuremath{\pi}\xspace}

 \def\Ptau        {\ensuremath{\tau}\xspace}

 \mathchardef\PDelta="7101
 \mathchardef\PXi="7104
 \mathchardef\PLambda="7103
 \mathchardef\PSigma="7106
 \mathchardef\POmega="710A
 \mathchardef\PUpsilon="7107
 \mathchardef\PPi="7105
                  
 \def\PB      {\ensuremath{B}\xspace}                 
                  
 \def\PD      {\ensuremath{D}\xspace}

 \def\PK      {\ensuremath{K}\xspace}

 \def\PW      {\ensuremath{W}\xspace}

 \def\PZ      {\ensuremath{Z}\xspace}                 
                  
 \def\Pb      {\ensuremath{b}\xspace}                 
 \def\Pc      {\ensuremath{c}\xspace}

 \def\Pi      {\ensuremath{i}\xspace}

 \def\Ps      {\ensuremath{s}\xspace}

 \def\thebaroffset{0.18em}
}
\newcommand{\offsetoverline}[2][\thebaroffset]{\kern #1\overline{\kern -#1 #2}}%

\makeatletter
\ifcase \@ptsize \relax
  \newcommand{\miniscule}{\@setfontsize\miniscule{4}{5}}
\or
  \newcommand{\miniscule}{\@setfontsize\miniscule{5}{6}}
\or
  \newcommand{\miniscule}{\@setfontsize\miniscule{5}{6}}
\fi
\makeatother

\DeclareRobustCommand{\optbar}[1]{\shortstack{{\miniscule (\rule[.5ex]{1.25em}{.18mm})}
  \\ [-.7ex] $#1$}}

\def\mup        {{\ensuremath{\Pmu^+}}\xspace}

\def\mumu       {{\ensuremath{\Pmu^+\Pmu^-}}\xspace}

\def\tautau     {{\ensuremath{\Ptau^+\Ptau^-}}\xspace}

\def\neu        {{\ensuremath{\Pnu}}\xspace}

\def\g      {{\ensuremath{\Pgamma}}\xspace}

\def\W      {{\ensuremath{\PW}}\xspace}
\def\Wp     {{\ensuremath{\PW^+}}\xspace}

\def\Z      {{\ensuremath{\PZ}}\xspace}

\def\squark    {{\ensuremath{\Ps}}\xspace}

\def\cquark    {{\ensuremath{\Pc}}\xspace}

\def\bquark    {{\ensuremath{\Pb}}\xspace}

\def\pion   {{\ensuremath{\Ppi}}\xspace}
\def\piz    {{\ensuremath{\pion^0}}\xspace}
\def\pip    {{\ensuremath{\pion^+}}\xspace}

\def\kaon    {{\ensuremath{\PK}}\xspace}

\def\KorKbar {\kern \thebaroffset\optbar{\kern -\thebaroffset \PK}{}\xspace}

\def\Kp      {{\ensuremath{\kaon^+}}\xspace}
\def\Km      {{\ensuremath{\kaon^-}}\xspace}

\def\Kstarz  {{\ensuremath{\kaon^{*0}}}\xspace}

\def\D       {{\ensuremath{\PD}}\xspace}

\def\DorDbar {\kern \thebaroffset\optbar{\kern -\thebaroffset \PD}\xspace}
\def\Dz      {{\ensuremath{\D^0}}\xspace}

\def\Dp      {{\ensuremath{\D^+}}\xspace}
\def\Dm      {{\ensuremath{\D^-}}\xspace}

\def\DpDm    {\ensuremath{\Dp {\kern -0.16em \Dm}}\xspace}

\def\Dstarp  {{\ensuremath{\D^{*+}}}\xspace}

\def\Dsp     {{\ensuremath{\D^+_\squark}}\xspace}

\def\B       {{\ensuremath{\PB}}\xspace}

\def\BorBbar {\kern \thebaroffset\optbar{\kern -\thebaroffset \PB}\xspace}

\def\Bd      {{\ensuremath{\B^0}}\xspace}

\def\BdorBdbar {\kern \thebaroffset\optbar{\kern -\thebaroffset \Bd}\xspace}

\def\Bs      {{\ensuremath{\B^0_\squark}}\xspace}

\def\BsorBsbar {\kern \thebaroffset\optbar{\kern -\thebaroffset \Bs}\xspace}

\def\Y#1S{\ensuremath{\PUpsilon{(#1S)}}\xspace}

\def\Lz          {{\ensuremath{\PLambda}}\xspace}

\def\LorLbar     {\kern \thebaroffset\optbar{\kern -\thebaroffset \PLambda}\xspace}

\def\Lc          {{\ensuremath{\Lz^+_\cquark}}\xspace}

\def\BF         {{\ensuremath{\mathcal{B}}}\xspace}
\def\BR         {\BF}

\newcommand{\decay}[2]{\ensuremath{#1\!\to #2}\xspace} 

\def\to                 {\ensuremath{\rightarrow}\xspace}

\def\BdKstGam     {\decay{\Bd}{\Kstarz \g}}

\def\AT#1     {\ensuremath{A_{\mathrm{T}}^{#1}}\xspace}           

\def\C#1      {\ensuremath{\mathcal{C}_{#1}}\xspace}                       
\def\Cp#1     {\ensuremath{\mathcal{C}_{#1}^{'}}\xspace}                    
\def\Ceff#1   {\ensuremath{\mathcal{C}_{#1}^{\mathrm{(eff)}}}\xspace}        
\def\Cpeff#1  {\ensuremath{\mathcal{C}_{#1}^{'\mathrm{(eff)}}}\xspace}       
\def\Ope#1    {\ensuremath{\mathcal{O}_{#1}}\xspace}                       
\def\Opep#1   {\ensuremath{\mathcal{O}_{#1}^{'}}\xspace}                    

\def\Dspg   {{\ensuremath{\Dsp}{\g}}\xspace}

\def\DzDzbg {{\ensuremath{\Dz}{\g}}\xspace}

\def\mpnu   {{\ensuremath{\mup}{\neu}}\xspace}

\def\DspFull{{\ensuremath{\Dsp\to\Kp\Km\pip}}\xspace}
\def\DzFull{{\ensuremath{\Dz\to\Km\pip}}\xspace}
\def\Dzkpipiz{{\ensuremath{\Dz\to\Km\pip\piz}}\xspace}

\def\WDspg   {\decay{\Wp}{\Dspg}}

\def\WDspgFull {\decay{\Wp}{{\Dsp{(\to\Kp\Km\pip)}}{\g}}\xspace}

\def\ZDzg    {\decay{\Z}{\DzDzbg}}
\def\ZDzgFull{\decay{\Z}{{\Dz{(\to\Km\pip)}}{\g}}\xspace}

\def\LcppKpi  {\decay{\Lc}{p\Km\pip}\xspace}

\def\DpKpipi {\decay{\Dp}{\Km\pip\pip}\xspace}

\def\DstarpDzpip {\decay{\Dstarp}{{\Dz(\to\Kp\Km)\pip}}}

\def\Zmm     {\decay{\Z}{\mumu}} 

\def\Wmpn    {\decay{\Wp}{\mpnu}}

\def\Ztt     {\decay{\Z}{\tautau}}

\def\ZDzDzbg {\decay{\Z}{\DzDzbg}}

\newcommand{\nospaceunit}[1]{\ensuremath{\text{#1}}}       
\newcommand{\aunit}[1]{\ensuremath{\text{\,#1}}}       

\newcommand{\tev}{\aunit{Te\kern -0.1em V}\xspace}
\newcommand{\gev}{\aunit{Ge\kern -0.1em V}\xspace}
\newcommand{\mev}{\aunit{Me\kern -0.1em V}\xspace}
\newcommand{\kev}{\aunit{ke\kern -0.1em V}\xspace}
\newcommand{\ev}{\aunit{e\kern -0.1em V}\xspace}
 
\newcommand{\mevc}{\ensuremath{\aunit{Me\kern -0.1em V\!/}c}\xspace}
\newcommand{\gevc}{\ensuremath{\aunit{Ge\kern -0.1em V\!/}c}\xspace}
\newcommand{\mevcc}{\ensuremath{\aunit{Me\kern -0.1em V\!/}c^2}\xspace}
\newcommand{\gevcc}{\ensuremath{\aunit{Ge\kern -0.1em V\!/}c^2}\xspace}

\def\mum  {\ensuremath{\,\upmu\nospaceunit{m}}\xspace}

\def\fb   {\ensuremath{\aunit{fb}}\xspace}
\def\invfb   {\ensuremath{\fb^{-1}}\xspace}

\def\ps   {\ensuremath{\aunit{ps}}\xspace}

\def\gsim{{~\raise.15em\hbox{$>$}\kern-.85em
          \lower.35em\hbox{$\sim$}~}\xspace}
\def\lsim{{~\raise.15em\hbox{$<$}\kern-.85em
          \lower.35em\hbox{$\sim$}~}\xspace}

\def\sqs   {\ensuremath{\protect\sqrt{s}}\xspace}

\def\pt         {\ensuremath{p_{\mathrm{T}}}\xspace}

\def\ptot       {\ensuremath{p}\xspace}
\def\et         {\ensuremath{E_{\mathrm{T}}}\xspace}

\def\mrad{\aunit{mrad}\xspace}

\def\evtgen     {\mbox{\textsc{EvtGen}}\xspace}

\def\geant      {\mbox{\textsc{Geant4}}\xspace}

\def\photos     {\mbox{\textsc{Photos}}\xspace}

\def\pythia     {\mbox{\textsc{Pythia}}\xspace}

\def\tell1  {TELL1\xspace}
\def\ukl1   {UKL1\xspace}

\newcommand{\ie}{\mbox{\itshape i.e.}\xspace}

\newcommand{\lhcborcid}[1]{\href{https://orcid.org/#1}{\hspace*{0.1em}\raisebox{-0.45ex}{\includegraphics[width=1em]{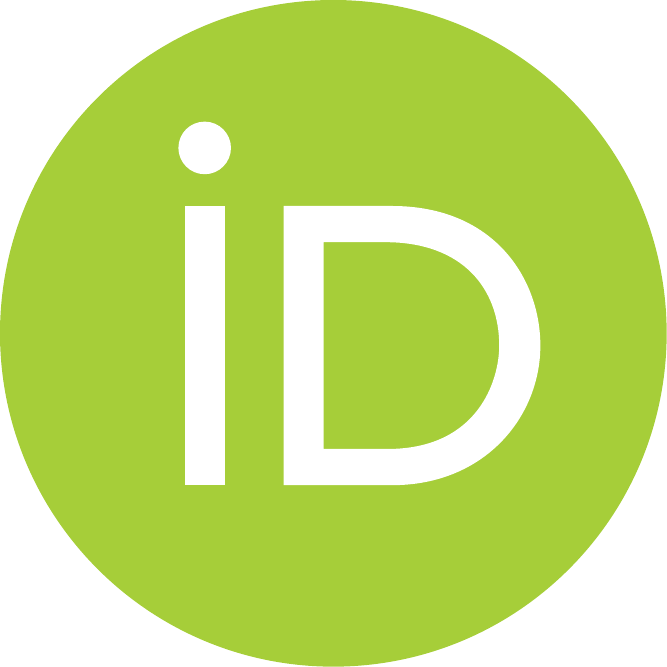}}}}

\usepackage{cite} 
\usepackage{mciteplus}

\usepackage{longtable} 
\usepackage{multirow}
\begin{document}
\renewcommand{\thefootnote}{\fnsymbol{footnote}}
\setcounter{footnote}{1}


\begin{titlepage}
\pagenumbering{roman}

\vspace*{-1.5cm}
\centerline{\large EUROPEAN ORGANIZATION FOR NUCLEAR RESEARCH (CERN)}
\vspace*{1.5cm}
\noindent
\begin{tabular*}{\linewidth}{lc@{\extracolsep{\fill}}r@{\extracolsep{0pt}}}
\ifthenelse{\boolean{pdflatex}}
{\vspace*{-1.5cm}\mbox{\!\!\!\includegraphics[width=.14\textwidth]{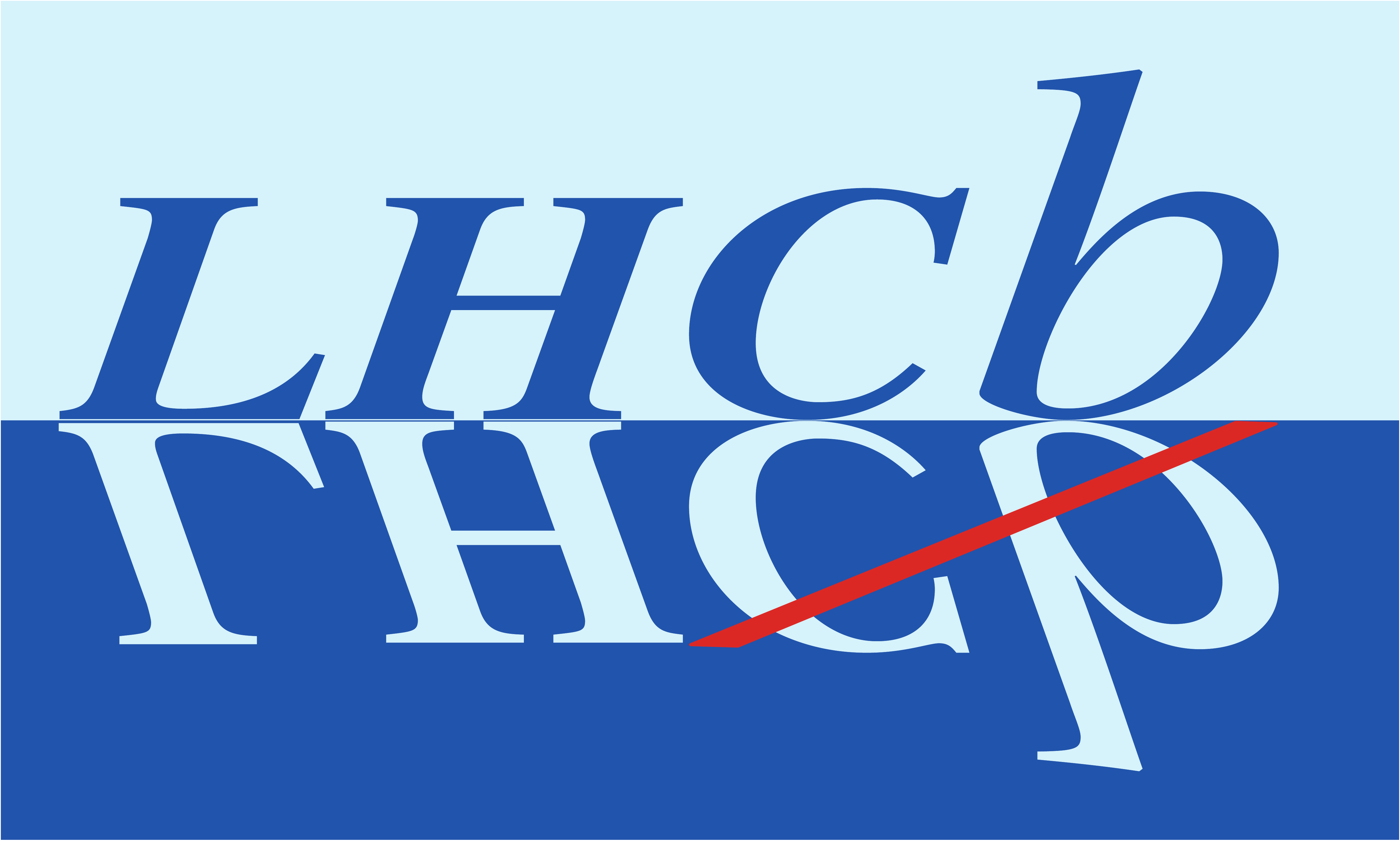}} & &}%
{\vspace*{-1.2cm}\mbox{\!\!\!\includegraphics[width=.12\textwidth]{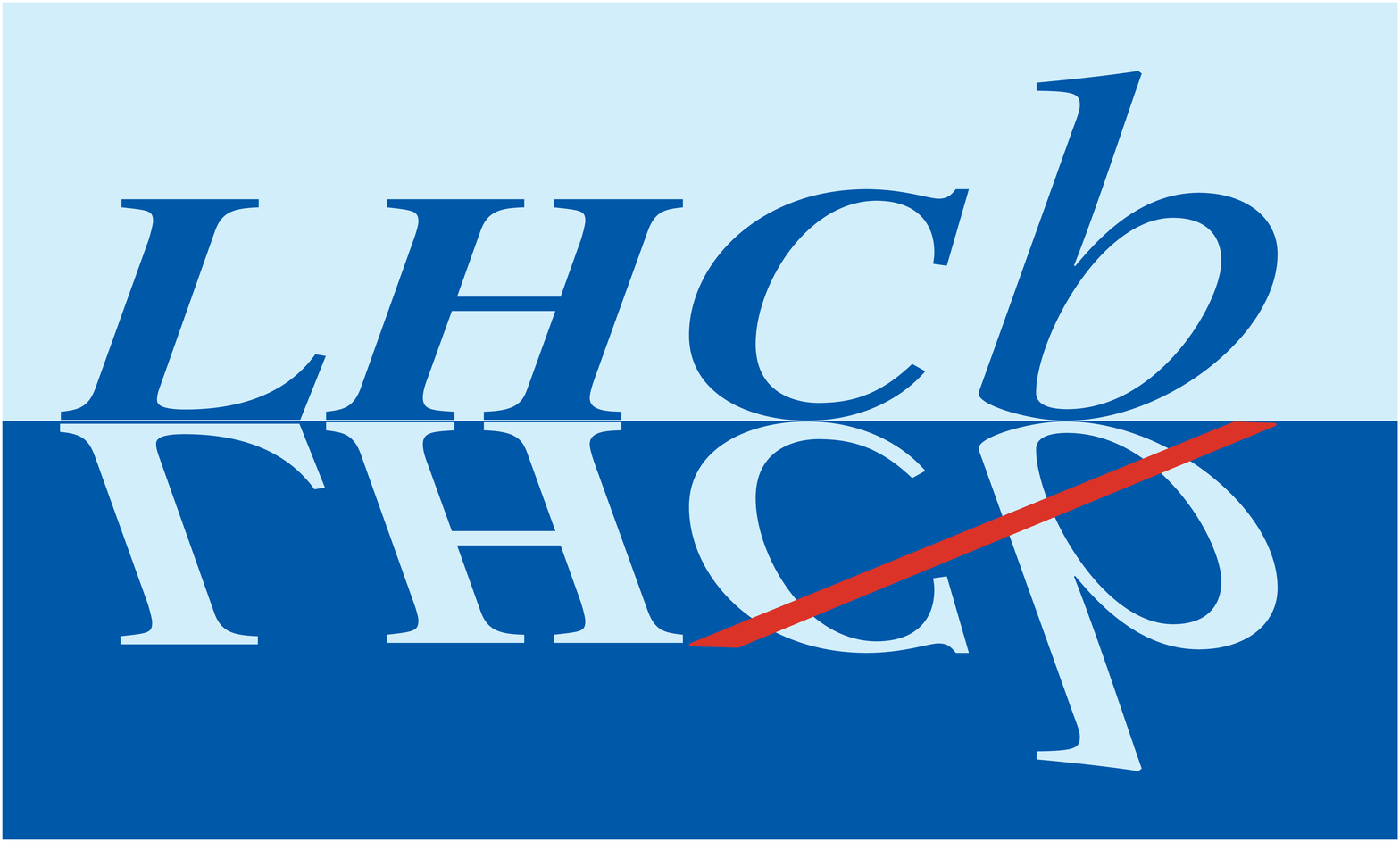}} & &}%
\\
 & & CERN-EP-2022-252 \\  
 & & LHCb-PAPER-2022-033 \\  
 & & \today \\ 
 & & \\
\end{tabular*}

\vspace*{4.0cm}

{\normalfont\bfseries\boldmath\huge
\begin{center}
  \papertitle 
\end{center}
}

\vspace*{2.0cm}

\begin{center}
\paperauthors 
\footnote{Authors are listed at the end of this paper.}
\end{center}

\vspace{\fill}

\begin{abstract}
  \noindent
 A search for the rare decays \WDspg and \ZDzDzbg is performed using proton-proton collision data 
 collected by the \lhcb experiment at a centre-of-mass energy of 13\tev, corresponding to an integrated luminosity of 2.0\invfb. 
 No significant signal is observed for either decay mode and upper limits on their branching fractions are set using \Wmpn and \Zmm decays as normalization channels.
 The upper limits are \WDsResults and \ZDzResults
 at 95\% confidence level 
 for the \WDspg and \ZDzg decay modes, respectively. 
  This is the first reported search for the \ZDzg decay,
  while the upper limit on the \WDspg branching fraction improves upon the previous best limit.

\end{abstract}

\vspace*{2.0cm}

\begin{center}
  Published in
  Chin.~Phys.~C47 (2023) 093002 
\end{center}

\vspace{\fill}

{\footnotesize 
\centerline{\copyright~\papercopyright. \href{\paperlicenceurl}{\paperlicence}.}}
\vspace*{2mm}

\end{titlepage}


\newpage
\setcounter{page}{2}
\mbox{~}


\renewcommand{\thefootnote}{\arabic{footnote}}
\setcounter{footnote}{0}

\cleardoublepage

\pagestyle{plain} 
\setcounter{page}{1}
\pagenumbering{arabic}


\section{Introduction}
\label{sec:Introduction}
The large production cross-sections of \W and \Z bosons
at hadron colliders offer unique opportunities to search for their rare decays, 
which can be used to test the Standard Model (SM) and probe for physics beyond the SM. 
Particularly interesting are the radiative decays of the \W and \Z bosons, 
predictions of their branching fractions using the quantum chromodynamics (QCD) factorization~\cite{Guberina:1980dc,Achasov:2012zzb,Huang:2014cxa,Arnellos:1981gy,Grossmann:2015lea,Manohar:1990hu} range from $10^{-6}$ to $10^{-12}$. 
Unlike in $B$-meson decays, where large power corrections to the decay rates
lead to sizable theoretical uncertainties, 
power corrections in \W and \Z boson decays are under good control due to the large energy
released to the final state hadrons. Thus, the study of hadronic-radiative \W and \Z bosons decays can provide stringent tests of the QCD factorization formalism~\cite{Grossmann:2015lea}.
After almost forty years from the discoveries of the \W and \Z bosons, 
no hadronic-radiative decay of these bosons has been observed, 
despite searches performed by the \atlas~\cite{ATLAS:2015vss,ATLAS:2016zln,ATLAS:2017gko,ATLAS:2018xfc,ATLAS:2022rej}, \cms~\cite{CMS:2018gcm}, and \cdf~\cite{Abe:1998vm} collaborations. 
The current best limit is $9\times 10^{-7}$ for the branching fraction of the $\Z \to \phi \gamma$ decay~\cite{ATLAS:2017gko}.

This paper presents searches for the radiative decays \WDspg
and \ZDzg, followed by \DspFull and \DzFull, at the \lhcb experiment, taking advantage of its excellent hadron identification capabilities.\footnote{The inclusion of charge-conjugate processes is implied throughout.}

\begin{figure}[!htpb]
\begin{center}
\includegraphics[height=0.22\textwidth]{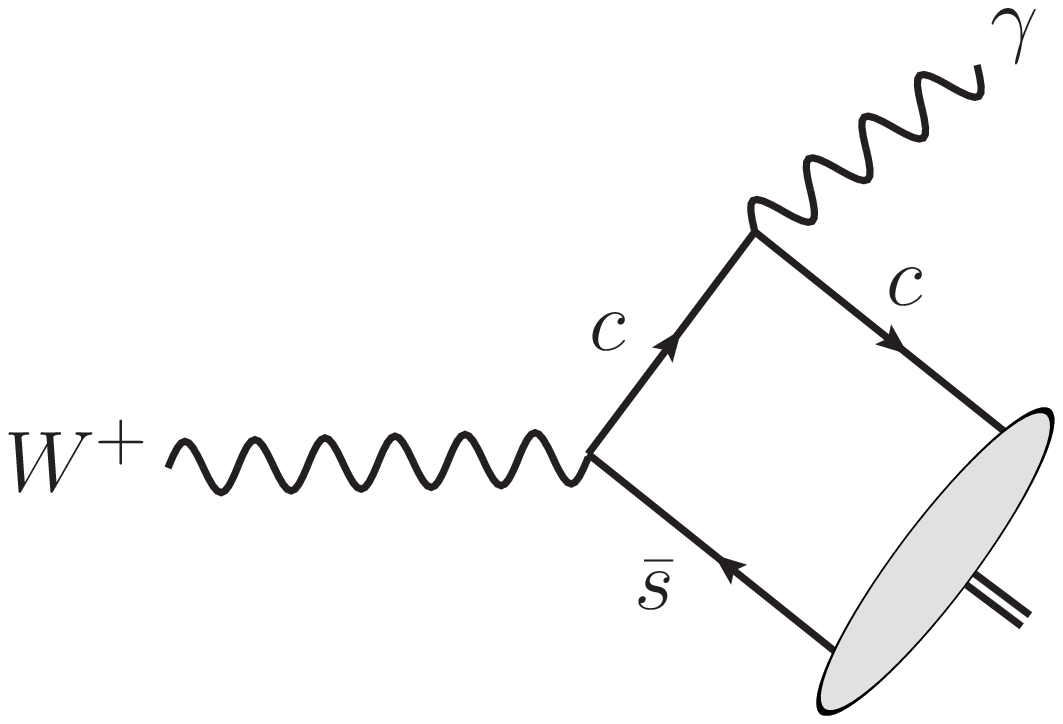}
\includegraphics[height=0.22\textwidth]{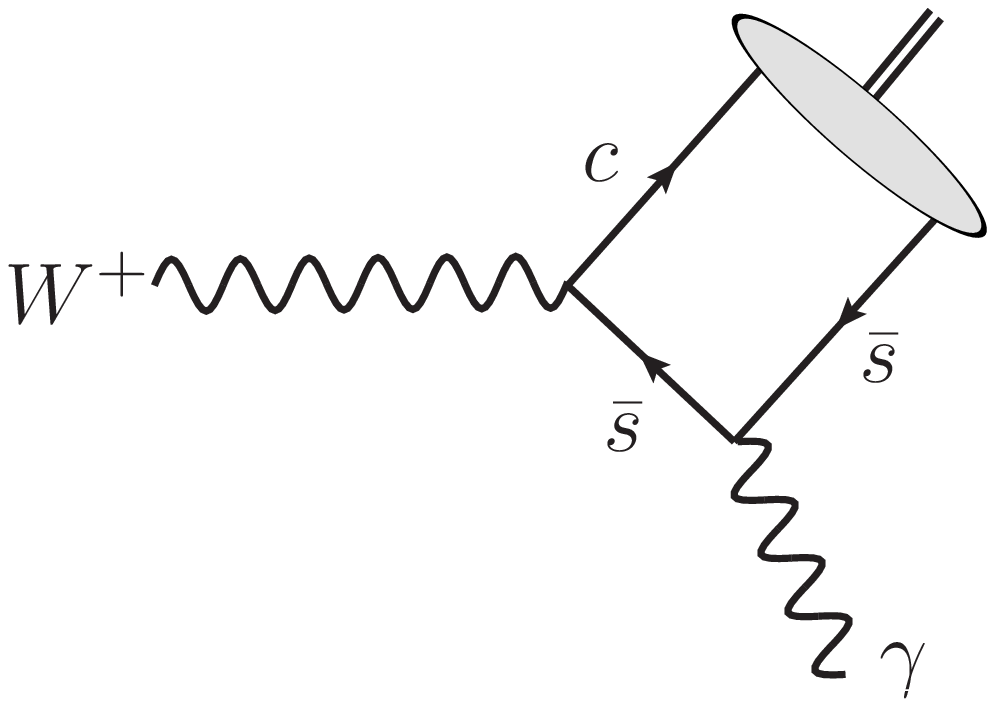}
\includegraphics[height=0.22\textwidth]{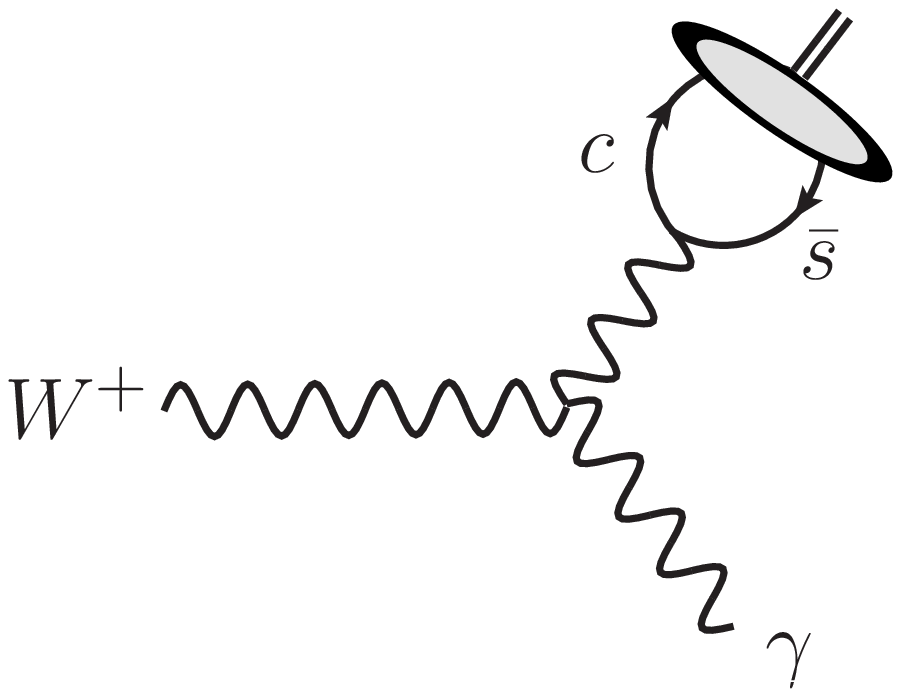}
\caption{Tree-level Feynman diagrams for the decay \WDspg.}
\label{fig:TheoryWBoson}
\end{center}
\end{figure}

The radiative decay \WDspg can proceed via tree-level diagrams as shown in
Fig.~\ref{fig:TheoryWBoson} and its branching fraction is predicted to be $(3.7\pm 1.5)\times 10^{-8}$ in the SM~\cite{Grossmann:2015lea}.
A previous search for this decay by the \cdf collaboration found no significant signal 
and determined the limit $\BR(\WDspg)<1.3 \times 10^{-3}$ at 95\% confidence level~\cite{Abe:1998vm}. 

\begin{figure}
\begin{center}
\includegraphics[width=0.35\textwidth]{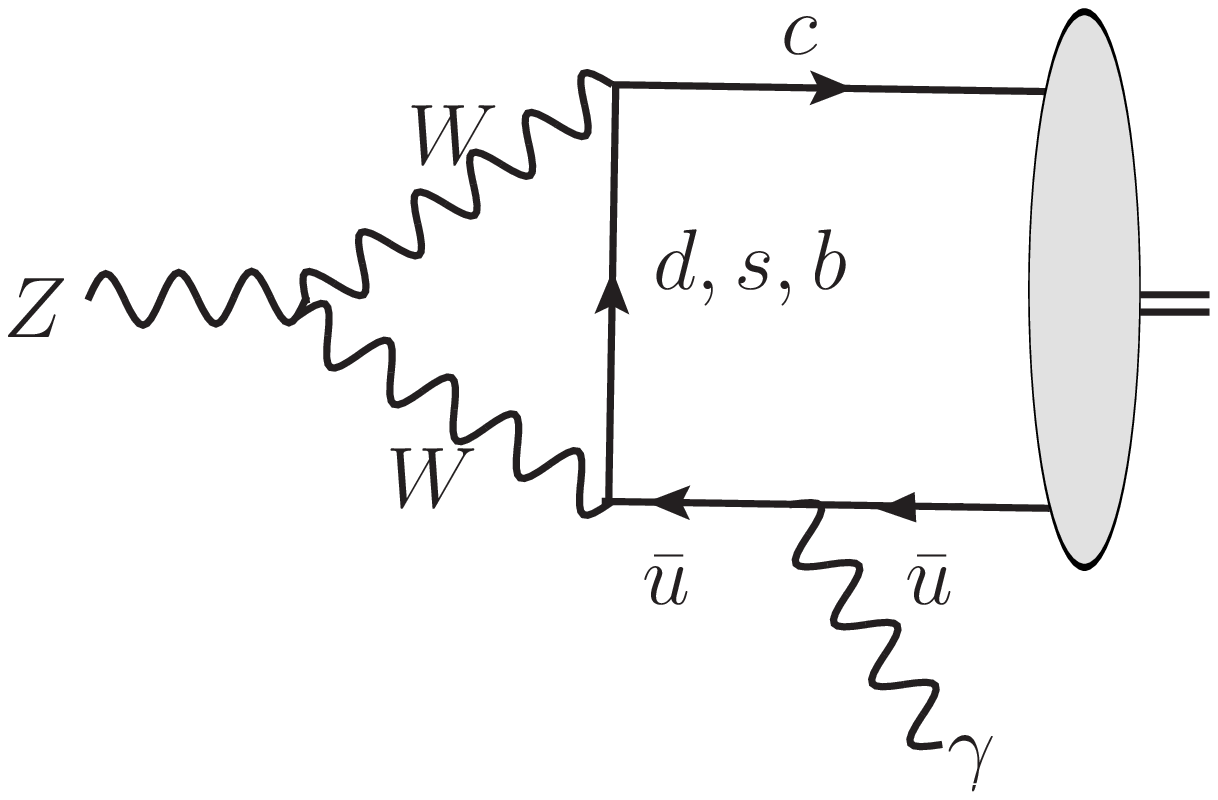}
\ \ \ \ \ 
\includegraphics[width=0.35\textwidth]{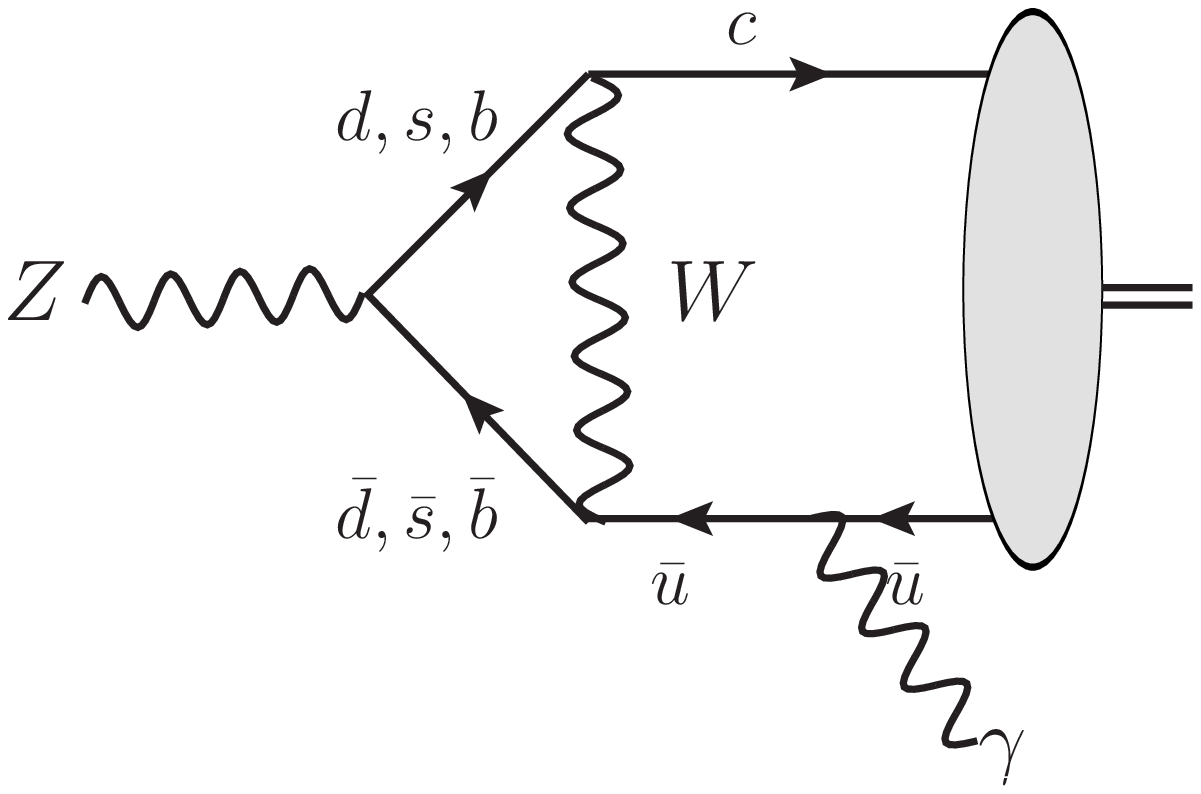}
\caption{Feynman diagrams contributing to the \ZDzg decay in the SM.}
\label{fig:TheoryZBoson}
\end{center}
\end{figure}

The radiative decay \ZDzg is a flavour-changing-neutral-current (FCNC) process, forbidden at the tree level, and can only proceed via higher-order loop processes in the SM, as shown in Fig.~\ref{fig:TheoryZBoson}.
The FCNC couplings of \Z boson are heavily constrained by the existing precision measurements from flavour physics, resulting in a
negligible branching fraction (${\cal{O}}\sim 10^{-15}$) of \ZDzg mode.
However, a search for this decay can provide a model-independent way to probe the FCNC couplings of the
$Z$ boson~\cite{Grossmann:2015lea}.
To date, there is no experimental study of this process. 

The search uses data from proton-proton collisions at 13\tev collected by the \lhcb experiment, corresponding to a luminosity of \lumi.
These data were collected in 2018 when the trigger selection dedicated to these decays was implemented.
Both relative and absolute branching fractions of the decays \WDspg and \ZDzg 
are determined, using \Wmpn and \Zmm decays as normalization channels, respectively. 
In this article, the fiducial criteria require
that the particle lie within the \lhcb acceptance, $10<\theta<400$\mrad for the charged particles, and $25<\theta_x<300$\mrad and $25<\theta_y<250$\mrad for the neutral particles, where $\theta$ is the polar angle with respect to the beam direction and $\theta_{x(y)}$ is the projection of the polar angle on the $xz$ ($yz$) plane.

\section{Detector and simulation}
The \lhcb detector~\cite{LHCb-DP-2008-001,LHCb-DP-2014-002} is a single-arm forward
spectrometer covering the \mbox{pseudorapidity} range $2<\eta <5$,
designed for the study of particles containing \bquark or \cquark
quarks. The detector includes a high-precision tracking system
consisting of a silicon-strip vertex detector surrounding the $pp$
interaction region~\cite{LHCb-DP-2014-001}, a large-area silicon-strip detector located
upstream of a dipole magnet with a bending power of about
$4{\mathrm{\,Tm}}$, and three stations of silicon-strip detectors and straw drift tubes~\cite{LHCb-DP-2013-003,LHCb-DP-2017-001} 
placed downstream of the magnet.
The tracking system provides a measurement of the momentum, \ptot, of charged particles with
a relative uncertainty that varies from 0.5\% at low momentum to 1.0\% at 200\gevc.
The minimum distance of a track to a primary $pp$ collision vertex (PV), the impact parameter (IP)~\cite{LHCb-DP-2014-002},
is measured with a resolution of $(15+29/\pt)\mum$,
where \pt is the component of the momentum transverse to the beam, in\,\gevc.
Different types of charged hadrons are distinguished using information
from two ring-imaging Cherenkov detectors~\cite{LHCb-DP-2012-003}.
Photons, electrons and hadrons are identified by a calorimeter system consisting of
scintillating-pad and preshower detectors, an electromagnetic
and a hadronic calorimeter~\cite{LHCb-DP-2014-002,LHCb-DP-2020-001}.
Charged and neutral clusters in the electromagnetic calorimeter are discerned by extrapolating the tracks reconstructed by the tracking system to the calorimeter plane. Photon and neutral pions are distinguished by cluster shape and energy distribution.
The read-out electronics of the electromagnetic calorimeter is optimised for the typical energy deposits that occur in \lhcb, \ie heavy flavour physics, which results in a maximum detectable transverse energy of 10\gev. 
A photon with very high energy is expected to have at least one saturated calorimeter cell.
Photons and neutral pions are distinguished by cluster shape, energy and mass distributions.
Muons are identified by a
system composed of alternating layers of iron and multiwire
proportional chambers~\cite{LHCb-DP-2012-002}.

The online event selection is performed by a trigger, which consists of a hardware stage followed by a two-level software stage. 
In between the two software stages, an alignment and calibration of the detector is performed in near real-time and their results are used in the trigger~\cite{LHCb-PROC-2015-011}.
The same alignment and calibration information is propagated to the offline reconstruction, ensuring a consistent and high quality reconstruction and event selection between the trigger and offline software. The identical performance of the online and offline 
reconstruction offers the opportunity to perform physics analyses directly using candidates reconstructed in the trigger~\cite{LHCb-DP-2012-004,LHCb-DP-2016-001}, 
which the present analysis exploits to reduce the event size by one order of magnitude. 

Simulation is required to model the effects of the detector acceptance and the imposed selection requirements.
In the simulation, $pp$ collisions are generated using \pythia~\cite{Sjostrand:2007gs} 
with a specific \lhcb configuration~\cite{LHCb-PROC-2010-056}. Decays of unstable particles
are described by \evtgen~\cite{Lange:2001uf}, in which final-state radiation is generated using \photos~\cite{davidson2015photos}.
The interaction of the generated particles with the detector, and its response, are implemented using the \geant toolkit~\cite{Allison:2006ve, *Agostinelli:2002hh} 
as described in Ref.~\cite{LHCb-PROC-2011-006}. 
In the signal simulation, the decay probability of \DspFull is taken to be uniform across the available phase space. 
However, there are sizable resonance contributions in the \Dsp decay, which are accounted for by assigning weights to the simulated candidates. The weights are evaluated by comparing the Dalitz distribution of the \Dsp decay~\cite{LHCb-PAPER-2018-033} and a phase-space model at the generator level.

\section{Reconstruction and selection}
The \WDspgFull and \ZDzgFull candidates 
must be matched to a positive decision in the hardware trigger,
which selects events by exploiting the particular signature of a photon with a high-energy cluster in the electromagnetic calorimeter 
and hadrons with high-transverse-energy deposits, $E_{\mathrm{T}}$, in the calorimeters.

In the first software trigger stage,
the charged final-state particles are required to 
be inconsistent with originating from a PV,
and are further required to pass a boosted decision tree (BDT) multivariate selector~\cite{Gligorov:2012qt}, 
where the input variables rely on transverse momentum, vertex fit quality, and flight distance information for \Dsp and \Dz candidates.
In the second software trigger stage, \DspFull candidates are reconstructed from three tracks each with $\pt > 0.25\gevc$. 
To further suppress background, additional requirements are applied to the combination of tracks, by requiring
at least one of the three tracks to have \pt greater than 1.0\gevc, at least two of them to have \pt greater than 0.4\gevc, and the scalar \pt sum of the three tracks to be greater than 3\gevc.
The \Dsp candidate is required to have a reconstructed invariant mass in the range $[1.879, 2.059]\gevcc$, 
and $\pt > 15\gevc$. The \Dsp candidate is further required to have a good vertex fit quality, 
and the \Dsp vertex is required to be displaced from every PV by a distance corresponding to a \Dsp decay time larger than 0.2\ps.
An additional neutral particle, which must be identified as a photon and have \et greater than 10\gev, is then combined with the \Dsp candidate
to form a \WDspgFull candidate. The \W candidate must have invariant mass between 30 and 130\gevcc.
Similarly, \DzFull candidates are reconstructed from two tracks that pass track quality requirements, with 
$\pt > 0.5\gevc$ and $p > 5\gevc$. 
The \Dz candidate is required to have reconstructed mass in the range $[1.715, 2.015]\gevcc$, 
and \pt greater than 15\gevc. 
The \Dz candidate must have a good vertex fit quality and must be displaced from every PV. 
Each \ZDzgFull candidate is reconstructed from a \Dz candidate plus a photon
with $\et > 10 \gev$.
The \Z boson candidates are required to have the invariant mass between 30 and 130\gevcc.

In the offline selection, the photon candidates are required to satisfy $\et > 15\gev$. Background due to photons from \piz decays is rejected by a dedicated algorithm~\cite{CalvoGomez:2015bwl}.
The final-state particles associated with the \DspFull candidates are further required to have \pt greater than 0.5\gevc and be located within the \lhcb fiducial region.
Two opposite-charged tracks are required to be identified as kaon tracks.
The \Dsp mass region is restricted to the range [1.92, 2.02]\gevcc,
while the \Dsp \pt is required to be greater than 20\gevc. 
A similar offline selection is applied to the \DzFull candidates, with the \Dz mass window requirement fixed to $[1.82, 1.92]\gevcc$.
Contamination from decays of other particles is suppressed by dedicated mass veto requirements.
The decay {$\DpKpipi$,} with a charged pion misidentified as a kaon, could contribute to the \DspFull decay.
Similarly the decay \DstarpDzpip
with the same final-state particles but different peak positions, could also contribute to the selected \Dsp candidates. 
A mass veto requirement $|M(\Km\pip\pip) - 1.870| > 0.02\gevcc$ is employed to reject the \Dp background. 
The background contribution from the \Dstarp 
process is removed with the condition $M(\Kp\Km) < 1.85\gevcc$.
The background contribution from \LcppKpi 
decay is studied and found to be well described as part of the combinatorial background. 
In total, 41\,718 and 73\,554 candidates are selected for \WDspg and \ZDzg, respectively.

The event selection of the normalization channels is similar 
to that used in previous \lhcb publications~\cite{LHCb-PAPER-2015-049,LHCb-PAPER-2021-037}. 
The online event selection is performed with the single muon triggers. At the hardware trigger stage, events are required to include a muon with high \pt.
The muon candidate must satisfy $\pt > 6\gevc$, $p > 8\gevc$, with a good track fit quality in the first software trigger stage. 
In the second software trigger stage, the muon candidate is further required to satisfy $\pt > 12.5\gevc$. 
For a \Wmpn candidate, the muon is required to pass all three single-muon trigger decision stages, and for a \Zmm candidate, at least one of the muons must pass all three decision stages.

The offline \Wmpn candidate selection starts by requiring events to have a well-reconstructed muon candidate~\cite{LHCb-DP-2014-002,Archilli_2013}
with \pt between 20 and 70\gevc and $2.0<\eta<4.5$.
The relative uncertainty in the momentum measurement for the muon is required to be less than 10\%. 
Different background contributions have been considered and specific selection conditions have been devised in order to suppress the various contributions. 
Since a muon from a \Wp boson decay is typically isolated from other particles, an isolation requirement is applied on the muon to suppress 
background contributions, where one muon is originating from a QCD process.
To reduce the contamination from \Zmm decays, 
events are required to have no additional muon candidate with \pt above 2\gevc.
Background contamination from \Ztt and heavy-flavour events is suppressed 
by requiring the IP 
of the muon candidate to be smaller than 40\mum. 
Muons from \W boson decay, which tend to be highly isolated, are associated with low energy deposits in the electromagnetic and hadronic calorimeters. 
Therefore, the amount of energy that is deposited in the calorimeters relative to the momentum of the track is required to be smaller than 4\%, 
which reduces the background from energetic pions and kaons punching through the calorimeters to the muon stations.
In total, 4.4 million \Wmpn candidates are selected with a purity of about 78\%. 

For the \Zmm offline selection, the candidates are required to have a pair of well-reconstructed tracks of opposite charge identified as muons. 
Muon tracks must have a \pt greater than 20\gevc and 2.0 $< \eta <$ 4.5.
The invariant mass of the two muons must be in the range of 60 to 120\gevcc. 
The relative uncertainty in the momentum measurement for each muon is required to be less than 10\%. 
In total, 320\,000 \Zmm candidates are selected with a purity of about 98.5\%.

\section{Yield determination}
To determine the yields of \Wp and \Z candidates, the pseudomass is used.
Similarly to the one used in Ref.~\cite{PseudoMass2021}, it is defined as
\begin{equation}
 m({M\gamma}) = \sqrt{2p^Mp_{\mathrm{T}}^M\frac{p^{\gamma}}{p_{\mathrm{T}}^{\gamma}}(1-\cos\theta)},
\end{equation}
where $p^M$ and $p_T^M$ are the momenta and transverse momenta of the meson, 
$p^{\gamma}$ and $p_T^{\gamma}$ are the momenta and transverse momenta of the photon, and $\theta$ is the opening angle between the meson and the photon.
The pseudomass is an approximation of the invariant mass in which the photon energy, which is poorly measured for transverse energies above the saturation value, is cancelled,
resulting in more than 5\% improvement on the expected upper limit.
The selected \WDspg candidates with pseudomass between 35 and 120\gevcc, and the selected \ZDzg candidates with pseudomass between 50 and 125\gevcc are used in the yield determination.
An extended maximum-likelihood method 
is adopted, with the finite template statistics accounted for, according to Ref.~\cite{Barlow:1993dm}. 
Upper limits on the signal yields are determined with the $CL_S$ method~\cite{cls,cls2}, using the candidate pseudomass and \pt distributions, and their correlations.
The upper limits are calculated at 95\% confidence level (C.L.), with the asymptotic $CL_S$ method in the 
\texttt{RooStats} framework~\cite{Moneta:2010pm,Cowan:2010js} taking into account systematic uncertainties. 

The signal shapes are determined from simulation after event selection.
The background shape is estimated using a background-dominated sample, following a data-driven method used previously by the \atlas collaboration~\cite{Chisholm:2021pdn,ATLAS:2022rej}.
The background-dominated data sample is selected using candidates in the meson invariant mass sideband, with the requirement on the meson \pt changed from 20\gevc to 15\gevc.
Since the \Dz lower mass sideband region contains background contributions from partially reconstructed decays (such as \DspFull with a missing \Kp and \Dzkpipiz with a missing \piz), the upper sideband region is selected $[1.91,2.00]$\gevcc for the \ZDzg background study. 
For \Dsp candidates, both lower $[1.90,1.94]$\gevcc and upper $[2.00,2.05]$\gevcc sideband regions are used. 
Probability density functions (PDFs) are used to model the distributions of the selected background events, 
and correlations between different variables are taken into account using Gaussian kernel density estimation (KDE)~\cite{Cranmer:2000du}. 
Pseudodata candidates are generated, from which the background shape in the 
discriminating variable is derived. 
The ensemble of pseudodata candidates is produced by randomly sampling distributions of the relevant kinematic variables.
These candidates are described by meson and photon four-momentum vectors: 
\begin{itemize}
 \item The meson four-momentum vector is constructed from its pseudorapidity ($\eta_M$), azimuthal angle ($\phi_M$), mass ($m_M$), and transverse momentum ($p_{\mathrm{T}}^M$). 
\item For the photon four-momentum vector, the $p_T^{\gamma}$ of the selected photon candidate is used, while the photon pseudorapidity, $\eta_\gamma$, and azimuthal angle, $\phi_\gamma$, are determined from the sampled $\Delta\eta(M,\gamma)$ and $\Delta\phi(M,\gamma)$ values, 
where $\Delta\eta(M,\gamma)$ and $\Delta\phi(M,\gamma)$ are the differences in $\eta$ and $\phi$ between the meson and the photon.
\end{itemize}
The correlations among these kinematic variables in background events are retained in the generation of the pseudodata through 
the following sampling scheme:
\begin{itemize}
\item The $\eta_M$, $\phi_M$, $m_M$, and $p_T^M$ values are drawn randomly and independently according to the corresponding PDFs using the meson sideband data events. 
In the background-dominated data samples, the correlations between these variables are found to be negligible, therefore the variables are assumed to be uncorrelated.
\item The distributions of $p_T^{\gamma}$, $\Delta\phi(M,\gamma)$ and $\Delta\eta(M,\gamma)$ are parameterised in bins of $p_T^M$, and values are drawn from the distributions of the bin corresponding to the previously generated $p_T^M$ value. 
\end{itemize}
Pseudodata candidates that pass the standard selection are used to construct two-dimensional template distributions of pseudomass and \pt.

\begin{figure}
\begin{center}
\includegraphics[width=0.45\textwidth]{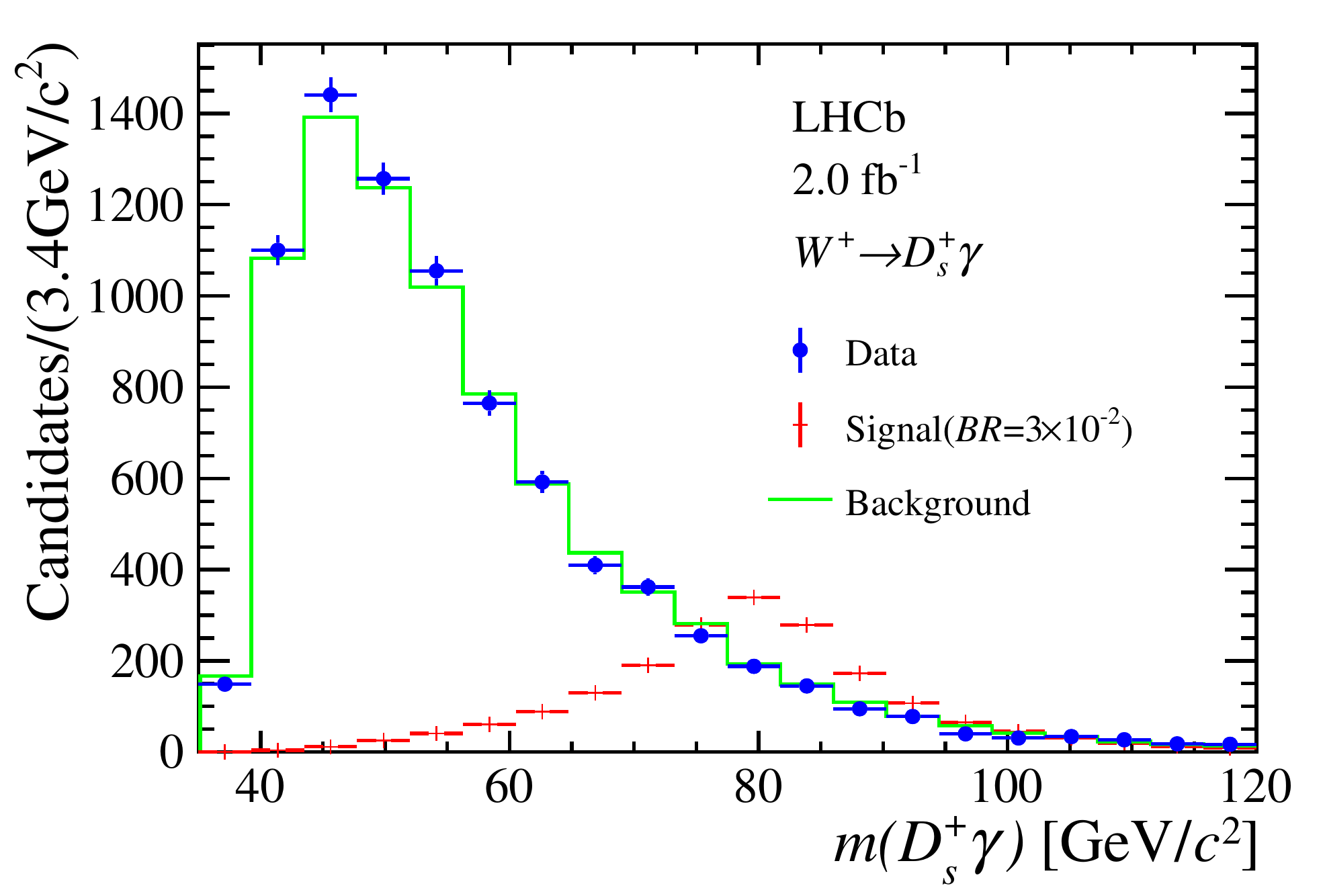}
\includegraphics[width=0.45\textwidth]{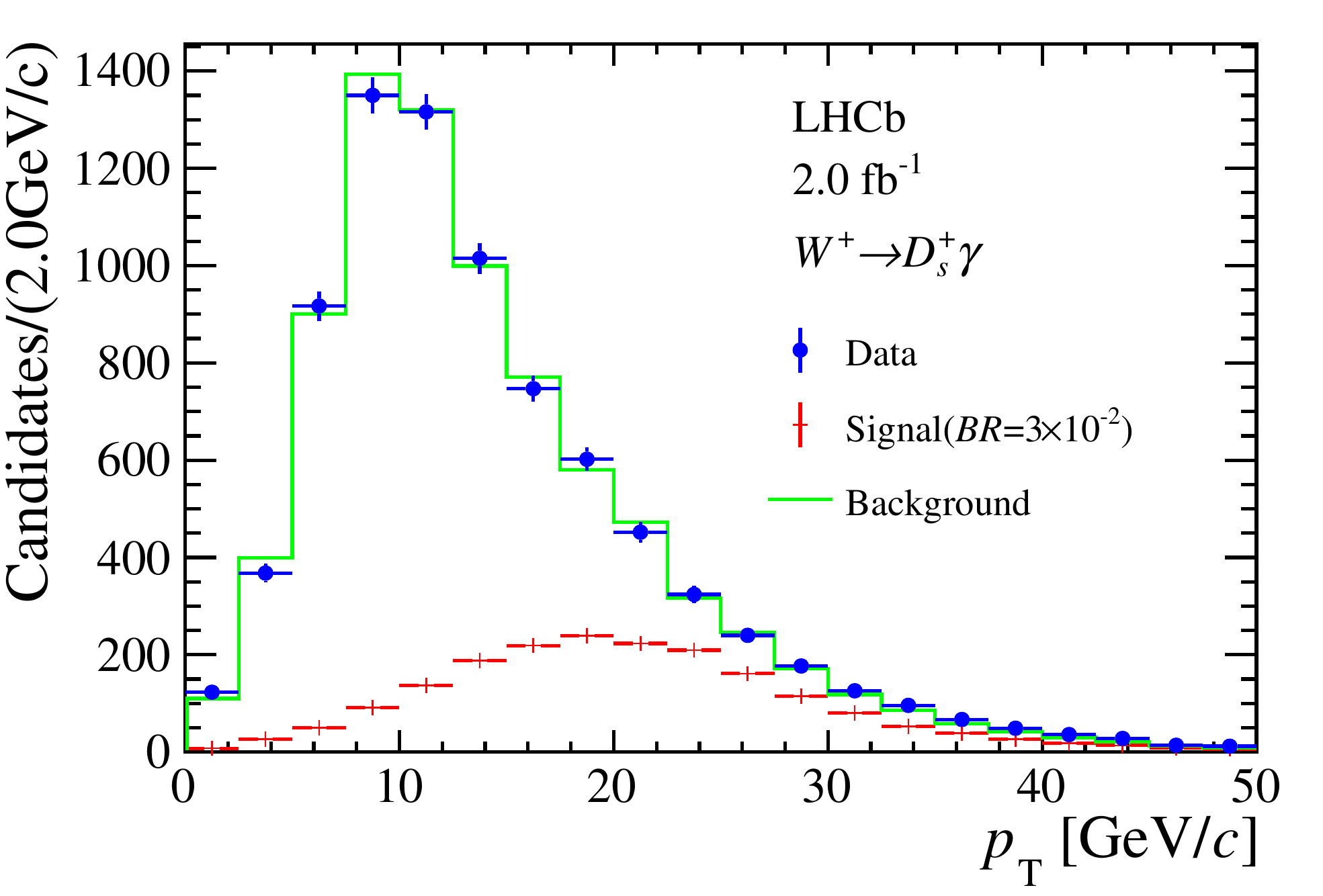}
\includegraphics[width=0.45\textwidth]{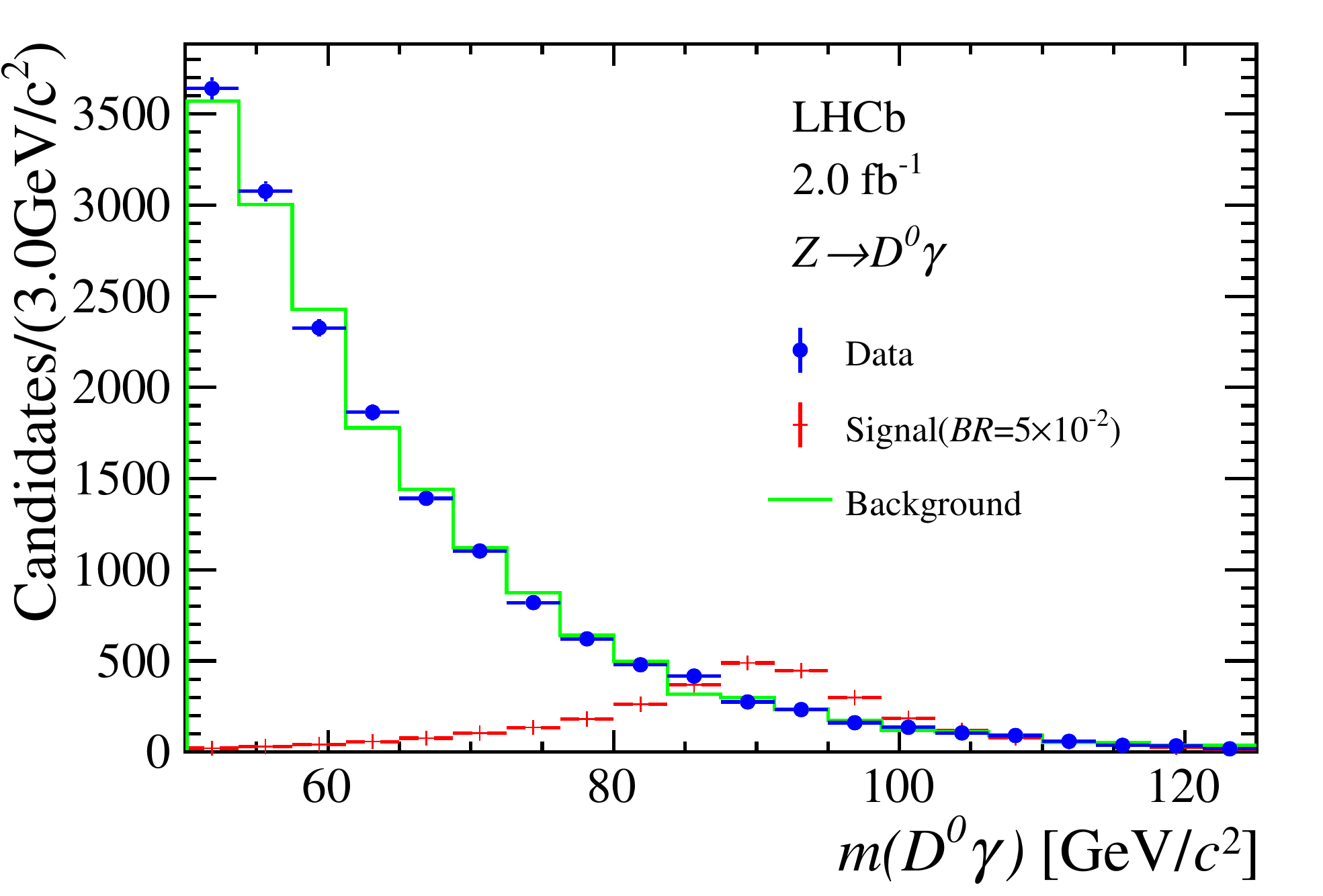}
\includegraphics[width=0.45\textwidth]{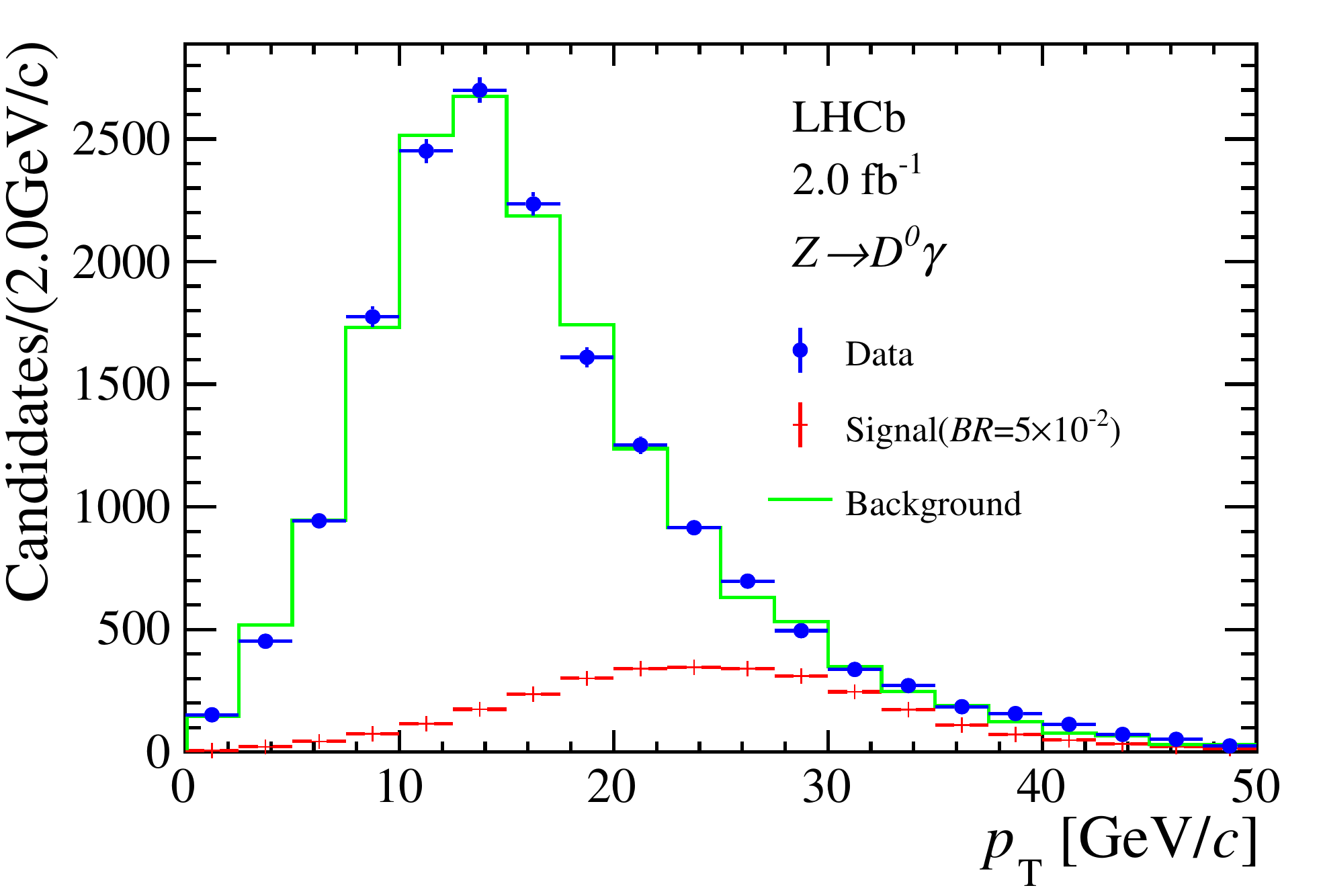}
\caption{Distributions of (left) pseudomass and (right) \pt for (upper) \WDspg and (lower) \ZDzg candidates. 
The blue points represent the selected data candidates, 
the red points represent simulated signal events, 
normalized to the branching fraction of \WDspg (\ZDzg) set to $3\times10^{-2}$ ($5\times10^{-2}$),
and the green lines represent the background shape, derived from pseudodata.}
\label{fig:MassFits_Boson}
\end{center}
\end{figure}

The distributions of pseudomass and \pt for the signal candidates, overlaid with the signal and background models, are shown in Fig.~\ref{fig:MassFits_Boson}. 
With no visible signal contribution, upper limits on the relative and absolute branching fractions are calculated.

\section{Ratio of branching fractions}
The ratio ${\cal{R}}(W)$ of the \WDspg branching fraction relative to that of the \Wmpn decay is defined as
\begin{eqnarray}
\label{eq:ratio_BR}
 {\cal{R}}(W) &\equiv& \frac{\BR(\WDspg)}{\BR(\Wmpn)} \nonumber \\
 &=& \frac{N_{s}\times \varepsilon_{n}\times {\cal{A}}_{n}}{N_{n}\times \varepsilon_{s} 
 \times 
 {\cal{A}}_{s}}\times \frac{1}{\BR(\DspFull)},
\end{eqnarray} 
where $\BR(\WDspg)$ is the branching fraction of \WDspg decay, 
$\BR(\Wmpn)$ is the branching fraction of $\Wmpn$ decay,
and $\BR(\DspFull)$ is the branching fraction of \DspFull process; $N_s$ ($N_n$) is the total signal (normalization) yield after event selection and background subtraction;
${\cal{A}}_{s}$ (${\cal{A}}_n$) is the 
probability for the true \W boson decay charged products momenta to lie within the \lhcb acceptance,
$\varepsilon_{s}$ ($\varepsilon_{n}$) is the total trigger, reconstruction and selection efficiency of the signal (normalization) channel.

The equivalent ratio of branching fractions, ${\cal{R}}(Z)$, and absolute branching fraction of the \ZDzg decay are studied using the \Zmm decay as a normalization channel. 
With the requirement that the final-state particle must be in the \lhcb detector fiducial region, the definition can be written as
\begin{eqnarray}
\label{eq:ratio_BRz}
 {\cal{R}}(Z) &\equiv& \frac{\BR{(\ZDzg)}}{\BR{(\Zmm)}} \nonumber \\
 &=& \frac{N_{s}\times \varepsilon_{n}\times {\cal{A}}_{n}}{N_{n}\times \varepsilon_{s} 
 \times 
 {\cal{A}}_{s}}\times \frac{1}{\BR(\DzFull)},
\end{eqnarray} 
where $\BR{(\ZDzg)}$ is the branching fraction of \ZDzg decay, 
$\BR(\Zmm)$ is the branching fraction of $\Zmm$ decay,
and $\BR(\DzFull)$ is the branching fraction of \DzFull process.

As the number of final-state particles is different between signal and normalization channels, 
the acceptance correction is determined and applied to the
$\cal{R}$ calculation.
The ${\cal{A}}_{s}$ and ${\cal{A}}_n$ factors are evaluated using event generators.
In the acceptance study, the uncertainty from parton distribution functions is taken as a systematic uncertainty.
The efficiencies $\varepsilon_{s}$ and $\varepsilon_{n}$ are determined from control and simulated samples.
For signal, the event selection efficiencies are determined from simulation, where
the track detection and particle identification efficiencies are calibrated with the data~\cite{LHCb-DP-2013-002,LHCb-PUB-2016-021,LHCb-DP-2018-001}.
The photon identification efficiency is calibrated using a \BdKstGam control sample which is widely tested and used for the calibration of photon variables inside the \lhcb collaboration, using only events with a photon of $\et > 10 \gev$.
The muon efficiencies are estimated using \Zmm data candidates with the tag-and-probe method~\cite{LHCb-PAPER-2021-037}. 

\section{Systematic uncertainties}
\label{sec:Systematic uncertainties}
The systematic uncertainties in the ${\cal{R}}(W)$ and ${\cal{R}}(Z)$ measurements are summarised in Table~\ref{tab:Tablesystemuncert}.
The uncertainties in the \DspFull and \DzFull branching fractions are 1.86\% and 0.76\%, respectively~\cite{PDG2022}.
The uncertainties of the normalization modes are expected to be uncorrelated with the uncertainties of the signal modes.
Systematic uncertainties from normalization channels are studied separately for the \Wmpn and \Zmm channels. 
Uncertainties from background estimation, efficiency calculations, signal determination and limited simulation sample size are taken 
into account, leading to relative uncertainties of 0.96\% for \Zmm and 3.08\% for \Wmpn decays in the ${\cal{R}}(W)$ and ${\cal{R}}(Z)$ measurements.
The \WDspg simulation is corrected using the measured Dalitz-plot distribution.
To determine the uncertainty from the meson decay modelling, the binning width of the reference Dalitz-plot distribution is varied by a factor of 0.75.

An uncertainty is assigned due to the limited size of the simulation samples used to determine the event selection efficiency.
The PID efficiency is calibrated using a control data sample~\cite{LHCb-DP-2020-001}, and a
systematic uncertainty arises due to the limited sample size. 
The uncertainty is estimated by enlarging or decreasing the binning of $p$, $\eta$, 
and event multiplicity of the control sample by a factor of two. 
Similarly, the systematic uncertainty associated with the photon identification efficiency calibration is 
evaluated by varying the binning of the \BdKstGam control sample.
The modelling of saturated calorimeter cells in the simulation is calibrated using \BdKstGam 
data events, and the systematic uncertainties from the size of the control sample and the binning schemes are studied.
The combined uncertainty is determined to be 3.0\% (3.1\%) for \ZDzg (\WDspg).
The uncertainty on the acceptance correction takes into account the size of the simulation samples, and uncertainties in the parton distribution function, factorisation and renormalisation scale dependencies as well as the uncertainty on $\alpha_s$, combined in quadrature. 

To estimate the systematic uncertainty from the background modelling, two sources are studied. 
The meson mass sideband regions are shifted to higher and lower masses by 0.03\gevcc, and the deviations are assigned as uncertainty.
An additional uncertainty is assigned 
by changing the binning of the meson \pt (by a factor of up to 50\%) in the nonparametric data-driven approach, using finer and coarser binnings. 
The uncertainty from background modelling is determined to be 0.08\% (0.36\%) for the ${\ZDzg}$ (${\WDspg}$) search.
The PV association algorithm was updated during the data-taking period, which causes a mismatch between \lhcb data and simulation, and introduces systematic effects in the efficiency estimation. 
A correction is studied and applied to the simulation, and an
uncertainty is assigned for this correction.
An additional correction is applied to the simulation to 
account for imperfect modelling of the resolution of the meson invariant mass,
by applying a 0.5\% (0.6\%) smearing correction to the \Dsp (\Dz) simulation.
A systematic uncertainty is evaluated by varying the resolution correction within its statistical uncertainty.

\begin{table}
\begin{center}
\caption{Relative systematic uncertainties (in \%) in the $\cal{R}$ measurements for the
\ZDzg and \WDspg decay modes. 
The total systematic uncertainty is obtained from
the sum in quadrature of all contributions.}
\begin{tabular}{lcc}
\hline
 Source & \ZDzg $(\%)$ & \WDspg $(\%)$ \\
\hline
 Meson BF             & 0.76 &  1.86 \\
 Normalization        & 0.96 &  3.08 \\
 Dalitz               &  -   &  0.24 \\
 MC sample size       & 0.11 &  0.09 \\
 PID                  & 0.09 &  0.17 \\
 Photon ID            & 2.32 &  0.95 \\
 Calorimeter saturation  & 3.00 &  3.10 \\  
 Background           & 0.08 &  0.36 \\
 Acceptance           & 0.57 &  0.82 \\ 
 PV association       & 0.57 &  0.29 \\ 
 Resolution           & 0.20 &  0.09 \\
 \hline 
 Total                & 4.07 &  4.94 \\
\hline
\end{tabular}
\label{tab:Tablesystemuncert}
\end{center}
\end{table}

\section{Results}
\label{sec:results}
\begin{figure}
\centering
\includegraphics[width=0.48\textwidth]{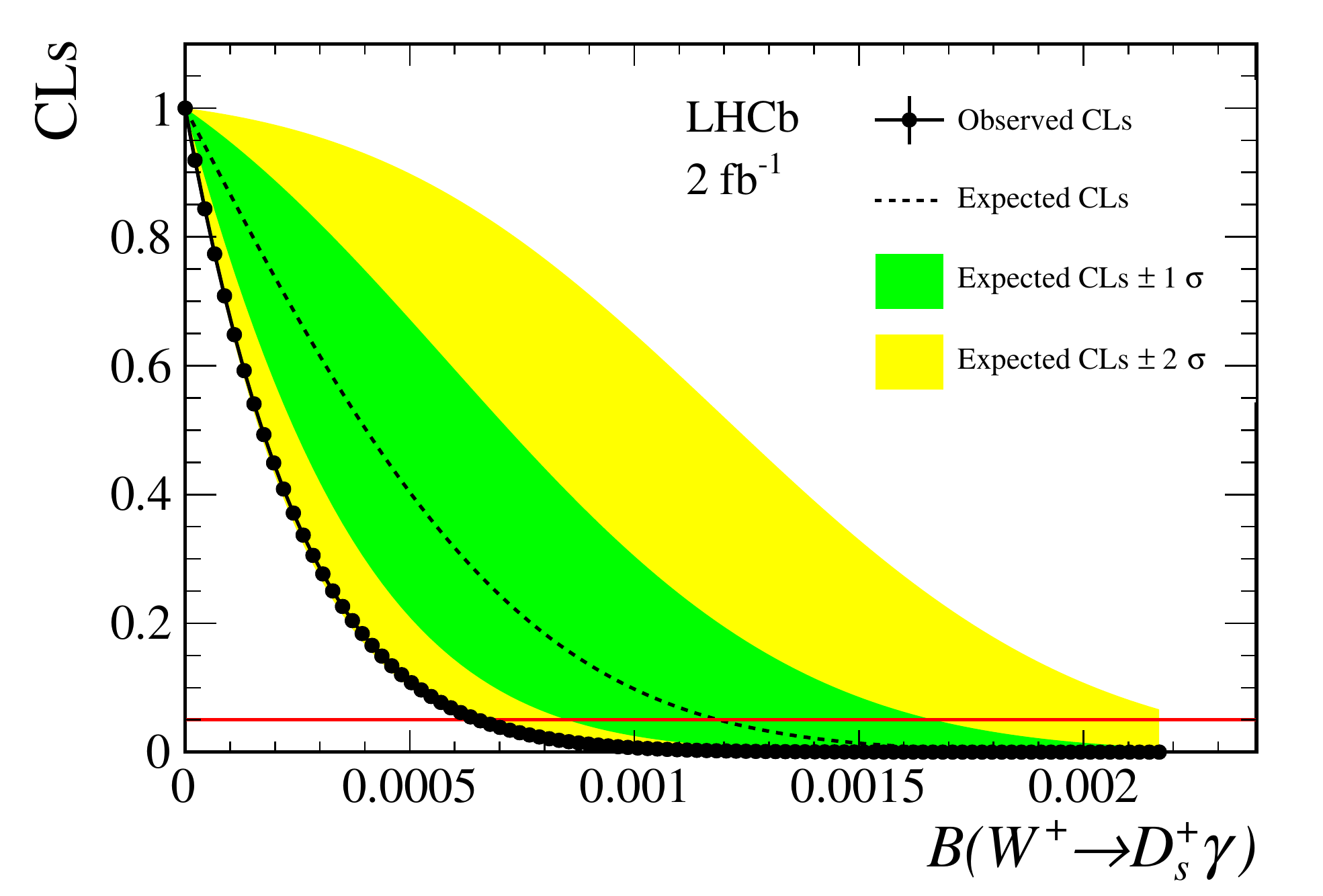}
\includegraphics[width=0.48\textwidth]{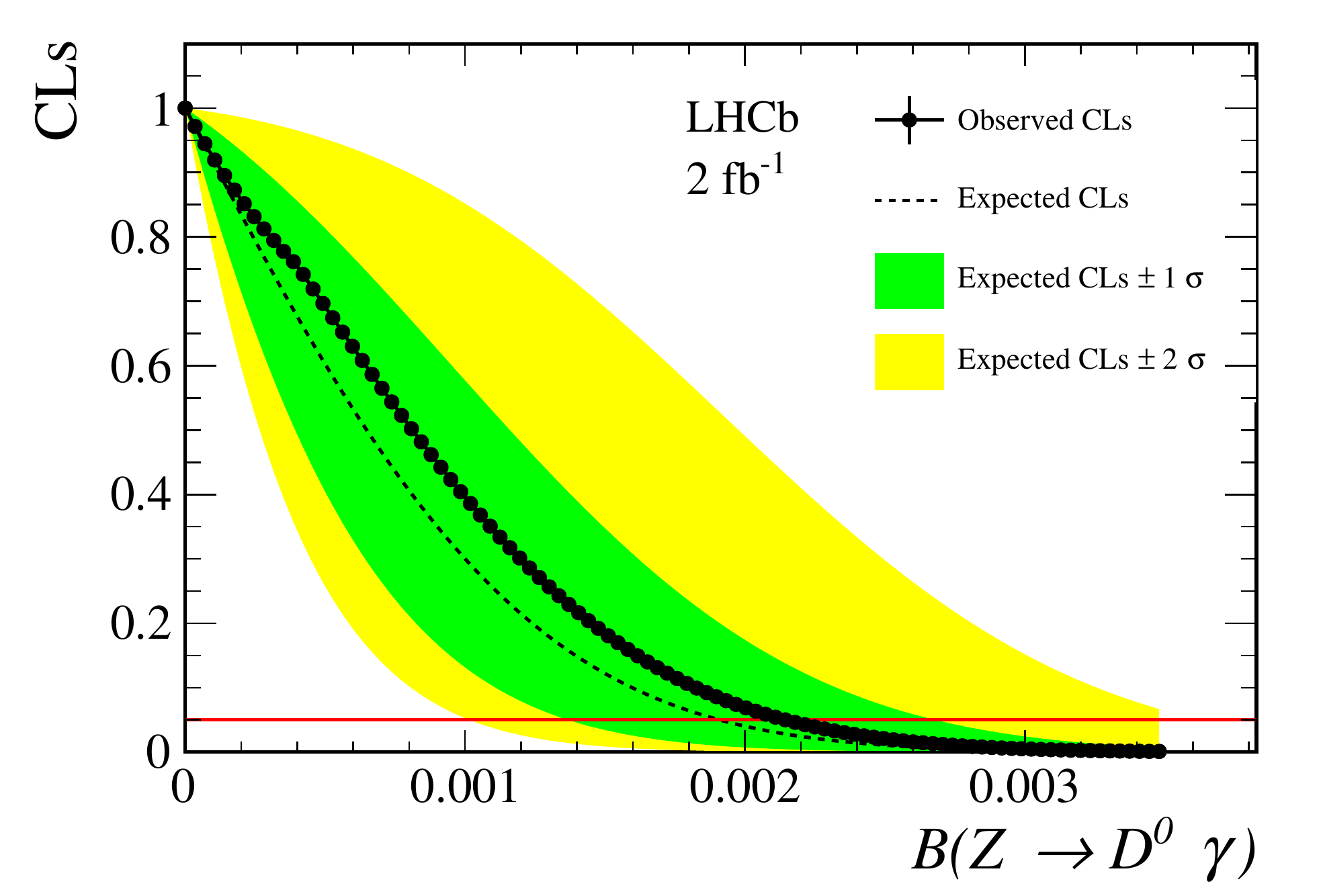}
\caption{Upper limits on branching fractions of the (left) \WDspg and (right) \ZDzg decays.}
\label{fig:ResultsWZ}
\end{figure}

No significant peaking structure is found in the inspected pseudomass ranges. The $CL_S$ method~\cite{cls,cls2} is used to calculate upper limits on the branching fractions of the \WDspg and \ZDzg decays.
In the calculation, we use the pseudomass and \pt distribution of $\Wp/Z$ boson candidates as observables, 
the signal shape is taken from the simulation after event selection, 
and the background distribution is estimated using a data-driven method.
The upper limit on the ratios of branching fractions are determined to be
\begin{eqnarray}
{{\cal{R}}(Z)} &<& \ZDzRelaResults \text{ at 95\% C.L.}, \nonumber \\
{{\cal{R}}(W)} &<& \WDsRelaResults \text{ at 95\% C.L.} \nonumber
\end{eqnarray}
The calculated and expected $CL_S$ exclusions are shown as a function of the branching fraction for \WDspg and \ZDzg decays in Fig.~\ref{fig:ResultsWZ}. 
The upper limits on the \WDspg and \ZDzg rare decay branching fractions are determined to be
\begin{eqnarray}
\BR(\ZDzg) &<& \ZDzResults \text{ at 95\% C.L.}, \nonumber \\
\BR(\WDspg) &<& \WDsResults \text{ at 95\% C.L.},\nonumber
\end{eqnarray}
using the known values~\cite{PDG2022} of the \Zmm and \Wmpn branching ratios. 
The expected upper limit on the branching fraction is calculated to be 
$1.2\times 10^{-3}$ ($1.9\times 10^{-3}$) for the \WDspg (\ZDzg) decay.

\section{Conclusion}
\label{sec:Conclusion}
Searches for the decays \WDspg and \ZDzg are performed using $pp$ collision data at $\sqs = 13$\tev collected by the \lhcb experiment in 2018, 
corresponding to an integrated luminosity of \lumi. 
No significant signal is observed above background.
For the \WDspg decay, the upper limit is determined to be \WDsResults at 95\% C.L., which is the best limit to date. 
The first search for the decay \ZDzg is performed, and the upper limit is determined to be \ZDzResults at 95\% C.L. 
These limits are well above the range of their SM predictions.
The next upgrade of the \lhcb detector~\cite{LHCb-PII-EoI} will allow operation at a much higher instantaneous luminosity 
($2 \times 10^{34}~cm^{-2} s^{-1}$), 
while keeping good performance in reconstructing and identifying hadrons of different species. This upgrade will be well-suited for further study of these very rare decays of the \W and \Z bosons.

\section*{Acknowledgements}
%
%
\noindent We express our gratitude to our colleagues in the CERN
accelerator departments for the excellent performance of the LHC. We
thank the technical and administrative staff at the LHCb
institutes.
We are indebted to the communities behind the multiple open-source software packages on which we depend.
We acknowledge support from CERN and from the national agencies:
CAPES, CNPq, FAPERJ and FINEP (Brazil); 
MOST and NSFC (China); 
CNRS/IN2P3 (France); 
BMBF, DFG and MPG (Germany); 
INFN (Italy); 
NWO (Netherlands); 
MNiSW and NCN (Poland); 
MEN/IFA (Romania); 
MICINN (Spain); 
SNSF and SER (Switzerland); 
NASU (Ukraine); 
STFC (United Kingdom); 
DOE NP and NSF (USA).
We acknowledge the computing resources that are provided by CERN, IN2P3
(France), KIT and DESY (Germany), INFN (Italy), SURF (Netherlands),
PIC (Spain), GridPP (United Kingdom), 
CSCS (Switzerland), IFIN-HH (Romania), CBPF (Brazil),
Polish WLCG  (Poland) and NERSC (USA).
Individual groups or members have received support from
ARC and ARDC (Australia);
Minciencias (Colombia);
AvH Foundation (Germany);
EPLANET, Marie Sk\l{}odowska-Curie Actions and ERC (European Union);
A*MIDEX, ANR, IPhU and Labex P2IO, and R\'{e}gion Auvergne-Rh\^{o}ne-Alpes (France);
Key Research Program of Frontier Sciences of CAS, CAS PIFI, CAS CCEPP, 
Fundamental Research Funds for the Central Universities, 
and Sci. \& Tech. Program of Guangzhou (China);
GVA, XuntaGal, GENCAT and Prog.~Atracci\'on Talento, CM (Spain);
SRC (Sweden);
the Leverhulme Trust, the Royal Society
 and UKRI (United Kingdom).

\addcontentsline{toc}{section}{References}
\bibliographystyle{LHCb}
\bibliography{main,LHCb-PAPER,LHCb-CONF,LHCb-DP,LHCb-TDR,myWZRare,standard}
 
\newpage

\centerline
{\large\bf LHCb collaboration}
\begin
{flushleft}
\small
R.~Aaij$^{32}$\lhcborcid{0000-0003-0533-1952},
A.S.W.~Abdelmotteleb$^{50}$\lhcborcid{0000-0001-7905-0542},
C.~Abellan~Beteta$^{44}$,
F.~Abudin{\'e}n$^{50}$\lhcborcid{0000-0002-6737-3528},
T.~Ackernley$^{54}$\lhcborcid{0000-0002-5951-3498},
B.~Adeva$^{40}$\lhcborcid{0000-0001-9756-3712},
M.~Adinolfi$^{48}$\lhcborcid{0000-0002-1326-1264},
P.~Adlarson$^{77}$\lhcborcid{0000-0001-6280-3851},
H.~Afsharnia$^{9}$,
C.~Agapopoulou$^{13}$\lhcborcid{0000-0002-2368-0147}
% ...

{\footnotesize \it

% ...
$^{9}$Universit{\'e} Clermont Auvergne, CNRS/IN2P3, LPC, Clermont-Ferrand, France\\
% ...
$^{13}$LPNHE, Sorbonne Universit{\'e}, Paris Diderot Sorbonne Paris Cit{\'e}, CNRS/IN2P3, Paris, France\\
% ...
$^{32}$Nikhef National Institute for Subatomic Physics, Amsterdam, Netherlands\\
% ...
$^{40}$Instituto Galego de F{\'\i}sica de Altas Enerx{\'\i}as (IGFAE), Universidade de Santiago de Compostela, Santiago de Compostela, Spain\\
% ...
$^{44}$Physik-Institut, Universit{\"a}t Z{\"u}rich, Z{\"u}rich, Switzerland\\
% ...
$^{48}$H.H. Wills Physics Laboratory, University of Bristol, Bristol, United Kingdom\\
% ...
$^{50}$Department of Physics, University of Warwick, Coventry, United Kingdom\\
% ...
$^{53}$School of Physics and Astronomy, University of Glasgow, Glasgow, United Kingdom\\
$^{54}$Oliver Lodge Laboratory, University of Liverpool, Liverpool, United Kingdom\\
% ...
$^{77}$Department of Physics and Astronomy, Uppsala University, Uppsala, Sweden, associated to $^{53}$\\
% ...
}
\end{flushleft}

\end{document}